\documentclass[twocolumn,secnumarabic,amssymb, nobibnotes, aps, prd]{revtex4-2}

\usepackage{amsmath, braket, graphicx, xcolor}
\graphicspath{{./images/}}

\begin{document}

\title{Squeezing Limit of the Josephson Ring Modulator\\as a Non-Degenerate Parametric Amplifier}%

\author{Dong Hwan Kim}%
\email[Dong Hwan Kim: ]{kiow639@add.re.kr}
\author{Su-Yong Lee}
\email[Su-Yong Lee: ]{suyong2@add.re.kr}
\author{Zaeill Kim}
\author{Taek Jeong}
\author{Duk Y. Kim}
\affiliation{Agency for Defense Development, Daejeon 34186, Korea}

\begin{abstract}
Two-mode squeezed vacuum states are a crucial component of quantum technologies. In the microwave domain, they can be produced by Josephson ring modulator which acts as a three-wave mixing non-degenerate parametric amplifier. Here, we solve the master equation of three bosonic modes describing the Josephson ring modulator with a novel numerical method to compute squeezing of output fields and gain at low signal power. We show that the third-order interaction from the three-wave mixing process intrinsically limits squeezing and reduces gain. Since our results are related to other general cavity-based three-wave mixing processes, these imply that any non-degenerate parametric amplifier will have an intrinsic squeezing limit in the output fields.
\end{abstract}
\maketitle

\textbf{Introduction.} Two-mode squeezed vacuum (TMSV) states are widely used in quantum technologies, such as continuous-variable quantum teleportation \cite{SB98, AF98},  quantum metrology \cite{PA10, SS13}, quantum dense coding \cite{SB00}, and quantum illumination \cite{ST08, CC19, SB20}. TMSV states, which consist of signal and idler modes, are usually generated via non-degenerate three-wave or four-wave mixing processes with the help of a strong pump beam \cite{AF98, ZO92, CM07, CE11, ME22}. Systems with such interactions are called as non-degenerate parametric amplifiers (NDPA). For applications in long distance target detection, preparation of TMSV states in the microwave range is beneficial. In microwave quantum technology, Josephson junctions provide nonlinearity for generation of squeezed states \cite{AB21, AR16}.

Ideal parametric amplifiers are systems with quadratic Hamiltonians, which are approximated via physical nonlinearities and external pumps. However, high-order terms of the Josephson junction potential have a negative effect on squeezing and gain in the degenerate parametric amplifier \cite{BK15, SB17}. For generation of TMSV states, we consider the NDPA, where the signal, idler, and pump modes each interact with different modes of a cavity. The Josephson ring modulator (JRM) structure acts as an NDPA, having advantages in that signal and idler modes are well separated in space and frequency \cite{NB10, NR12}. The effect of high-order terms of JRM on saturation power was studied in \cite{CL20} using a semi-classical approach. To identify the effect of actual JRM Hamiltonian on squeezing, we perform a quantum analysis of the NDPA master equation with three interacting bosonic modes in frequency domain, which is a computationally intensive task due to the massive size of the density matrix. Previously, stochastic methods were studied to bypass this problem in time domain under other systems \cite{KM93, YC95}, whereas obtaining frequency domain information from these methods is not straightforward. We instead develop a numerical method to solve the master equation on commercial computers and compute squeezing of the JRM output field along with gain at low signal power. We observe a limit in squeezing and find the fundamental reason by examining several different Hamiltonians.

\begin{figure}[t]
\includegraphics[width=8.5cm]{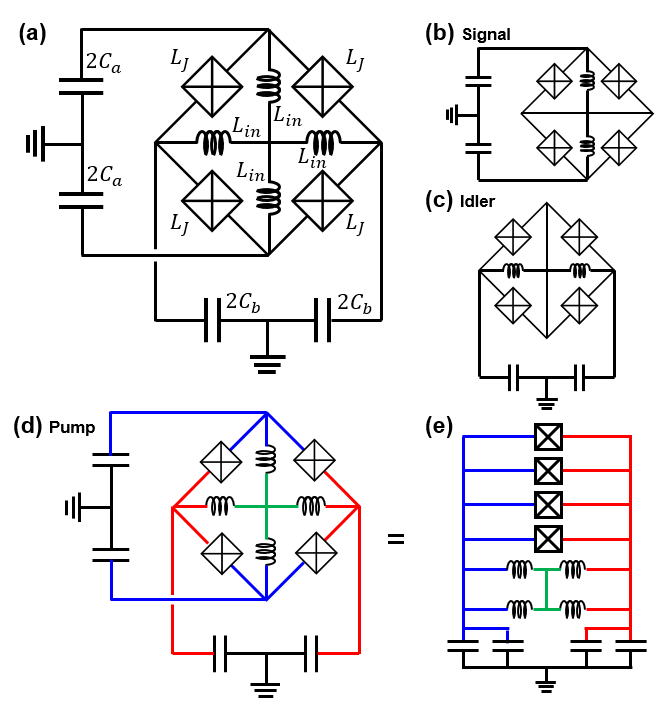}
\caption{(a) Circuit structure of JRM. Four identical Josephson junctions form a ring, connected to external capacitors. Internal inductors are added to operate at the Kerr nulling point. (b), (c), (d) LC circuit structure of the three resonance modes, signal, idler, pump respectively. (e) Equivalent circuit of (d), to clarify the LC circuit structure.}
\label{fig:cir}
\end{figure}

\textbf{JRM Hamiltonian and Approximations.} JRM consists of four identical Josephson junctions in a ring connected to external capacitors to form a microwave cavity as shown in Fig. \ref{fig:cir}(a). The junction is characterized by its critical current $i_c$. The Josephson energy and Josephson inductance are $E_J = \hbar \omega_J = \phi_0 i_c$ and $L_J = \phi_0 / i_c$ respectively, where $\phi_0 = \hbar/2e$ is the flux quantum. By adding additional internal inductors $L_{in}$, it is possible to operate JRM at the Kerr nulling point, where all even-order interactions are zero \cite{CL20}. The ratio of inductances $\beta = L_J/L_{in}$ controls the overall strength of nonlinearity.

JRM has three resonance modes, $\hat{a}$, $\hat{b}$, and $\hat{c}$, which we refer as signal, idler, and pump modes respectively. The Hamiltonian of JRM with an external pump on mode $\hat{c}$ is
\begin{align} \begin{split} \label{JRM}
H_{JRM}& = \sum_{m = a,b,c} \omega_m \hat{m}^\dagger \hat{m} + i\epsilon (\hat{c}^\dagger e^{-i\omega_P t} - \hat{c} e^{i\omega_P t} ) \\
& \hspace{10mm}  - 4\omega_J \sin \frac{\hat{\varphi}_a}{2} \sin \frac{\hat{\varphi}_b}{2} \sin \hat{\varphi}_c ,
\end{split} \end{align}
where $\omega_a$, $\omega_b$, and $\omega_c$ are resonance frequencies of the cavity. The external pump frequency is $\omega_P \simeq \omega_a + \omega_b$, and $\epsilon$ describes the external pump on mode $\hat{c}$. The resonance frequencies are determined by the (linearized) LC circuits in Fig. \ref{fig:cir} (b), (c), (e). The mode fluxes $\hat{\varphi}_{m}$ are related to quadrature operators $\hat{x}_m = \frac{1}{\sqrt{2}} (\hat{m} + \hat{m}^\dagger)$ ($m = a,b,c$) as \cite{SUPP}
\begin{equation} \label{mode flux}
\hat{\varphi}_a = \sqrt{\frac{2 \omega_a}{\beta \omega_J}}\hat{x}_a,\, \hat{\varphi}_b = \sqrt{\frac{2 \omega_b}{\beta \omega_J}}\hat{x}_b,\, \hat{\varphi}_c = \sqrt{\frac{ \omega_c}{\beta \omega_J}}\hat{x}_c .
\end{equation}

The linear approximation of $\sin \hat{\varphi} \simeq \hat{\varphi}$ with rotating wave approximation (RWA) gives the third-order three-wave mixing Hamiltonian. In a frame rotating at input field frequencies $\omega_S$, $\omega_I$, and $\omega_P$ for modes $\hat{a}$, $\hat{b}$, and $\hat{c}$ respectively, the approximation of Eq. (\ref{JRM}) is
\begin{equation} \label{3SoP}
H_{3} = \sum_{m = a,b,c} \Delta_m \hat{m}^\dagger \hat{m} + i\epsilon \hat{c}^\dagger  + i\sqrt{\frac{\omega_a \omega_b \omega_c}{2\beta^3 \omega_J}}\hat{a} \hat{b} \hat{c}^\dagger + \text{c.c.},
\end{equation}
where $\Delta_a = \omega_a - \omega_S$, $\Delta_b = \omega_b - \omega_I$, and $\Delta_c = \omega_c - \omega_P$ are detunings between cavity resonances and input fields. The phase of each mode is adjusted for the prefactor $i$ of interaction. The fifth-order approximation $H_5$ is obtained by adding terms from $\hat{\varphi}_a^3 \hat{\varphi}_b \hat{\varphi}_c$, $\hat{\varphi}_a \hat{\varphi}_b^3 \hat{\varphi}_c$, and $\hat{\varphi}_a \hat{\varphi}_b \hat{\varphi}_c^3$ in the sine power series to $H_3$. The stiff pump approximation Hamiltonians $H_{3s}$ and $H_{5s}$, where back-action of signal and idler modes on pump is neglected, arise from replacing $\hat{c}$ with $\epsilon /(\frac{\kappa_c}{2} +i\Delta_c)$ in $H_3$ and $H_5$. \cite{AK09}. The third-order stiff pump approximation Hamiltonian is the NDPA model Hamiltonian.
\begin{align}
H_{3s}=H_{NDPA} &= \sum_{m=a,b} \Delta_m \hat{m}^\dagger \hat{m} + i \left(g \hat{a} \hat{b} - g^* \hat{a}^\dagger \hat{b}^\dagger \right),\label{ham:NDPA}\\
g&= \sqrt{\frac{\omega_a \omega_b \omega_c}{2\beta^3 \omega_J}} \frac{\epsilon }{(\frac{\kappa_c}{2} -i\Delta_c)}.
\end{align}
For the system of Eq. (\ref{ham:NDPA}), squeezing and gain both increase unboundedly as pump power increases below a threshold power. Above this threshold, the system becomes unstable.

The whole JRM Hamiltonian is described by matrix elements of $\sin{\alpha \hat{x}}$ which are a sum of matrix elements of displacement operators \cite{KC69}. In general, $\braket{n|\sin\alpha\hat{x}|m}$ is nonzero when $n$ and $m$ have different parity. For small $\alpha$, the matrix elements $\braket{n|\sin\alpha\hat{x}|n\pm1}$ are dominant among all matrix elements \cite{SUPP}, hence we approximate the sine potential with only transitions like $\hat{a} \hat{b} \hat{c}^\dagger$ or $\hat{a}^\dagger \hat{b}^\dagger \hat{c}$, denoted as $H_{J1}$. Applying RWA, the next contribution comes from transitions like $\hat{a}^3 \hat{b}^3 \hat{c}^{\dagger 3}$ which have smaller matrix elements than $\hat{a} \hat{b} \hat{c}^\dagger$ like transitions. It was verified that considering these transitions does not affect the results obtained from $H_{J1}$. Explicit formulas for the Hamiltonians are given in \cite{SUPP}.

\textbf{Numerical Methods.} Squeezing $S$ is computed from output field moments as follows:
\begin{align}
S &= 10\, \text{Log}_{10} \frac{2}{\Delta},\; \Delta = \text{Var} ( \hat{X}_{-} ) + \text{Var} (\hat{P}_{+}),\\
\hat{X}_- &= \tfrac{1}{\sqrt{2}} \left(\hat{a}_{out} + \hat{a}_{out}^\dagger - e^{i\phi} \hat{b}_{out} - e^{-i\phi} \hat{b}_{out}^\dagger \right),\\
\hat{P}_+ &= \tfrac{1}{i\sqrt{2}} \left(\hat{a}_{out} - \hat{a}_{out}^\dagger + e^{i\phi} \hat{b}_{out} - e^{-i\phi} \hat{b}_{out}^\dagger \right).
\end{align}
$\hat{a}_{out}$ and $\hat{b}_{out}$ are output operators of signal and idler modes respectively, and $\phi$ is defined to minimize $\Delta$. It is known that $\Delta \geq 2$ for separable states \cite{RS00, LD00}. For a three-wave mixing process, the minimum of $\Delta$ is
\begin{equation}
\frac{\Delta_{min}}{2} = \braket{\hat{a}_{out}^\dagger \hat{a}_{out}} + \braket{\hat{b}_{out}^\dagger \hat{b}_{out}} +1 -2 \left| \braket{\hat{a}_{out} \hat{b}_{out}} \right|.
\end{equation}

It is possible to compute output field moments from the master equation and input-output relations \cite{CG85, SUPP},
\begin{align}
\dot{\rho} &= -i[H, \rho] + \sum_{m=a,b,c} \frac{\kappa_m}{2} \mathcal{D}_{\hat{m}} [\rho] =:\mathcal{L}[\rho], \label{me}\\
\hat{m}_{out} &= \sqrt{\kappa_m} \hat{m} - \hat{m}_{in} \hspace{3mm} (m=a,b,c), \label{bc}
 \end{align}
where $\hat{m}_{in(out)}$ are input (output) operators of each mode. $\mathcal{D}_{\hat{m}} [\rho] := 2 \hat{m} \rho \hat{m}^\dagger - \rho \hat{m}^\dagger \hat{m} - \hat{m}^\dagger \hat{m} \rho$ gives dissipation and $\kappa_m$ is the mode decay rate. For calculating the moments in frequency domain, the steady state $\rho_{ss}$ of the master equation and expressions like $\mathcal{L}^{-1} [\hat{a} \rho_{ss}]$ are needed. Both quantities are obtained by solving the master equations $\mathcal{L}[\rho_{ss}] = 0$ and $\mathcal{L}[\tilde{\rho}] = \hat{a} \rho_{ss}$. This is a nontrivial task, because there are $O(n^4 n_c^2)$ variables describing the density matrix when considering $n$ levels for the signal and idler modes each and $n_c$ levels for the pump mode. However, the master equations of three-wave mixing processes are block diagonal where blocks are defined by $a_1-b_1-a_2+b_2 = \text{(const.)}$, $\rho = \sum \rho_{a_1 b_1 c_1}^{ a_2 b_2 c_2} \ket{a_1, b_1, c_1}\bra{a_2, b_2, c_2}$, so we only need to solve on each block. We denote the subspace with $a_1-b_1-a_2+b_2 = k$ as $V_k$ and refer to variables in such a subspace as \textit{reduced variables}. The steady state is concentrated on $V_0$, and the number of variables in $V_0$ is $O( n^3  n_c^2)$. After reducing variables, the problem requires a reasonable amount of memory allowing us to solve the problem.

\begin{figure*}[t]
\includegraphics[width=18cm]{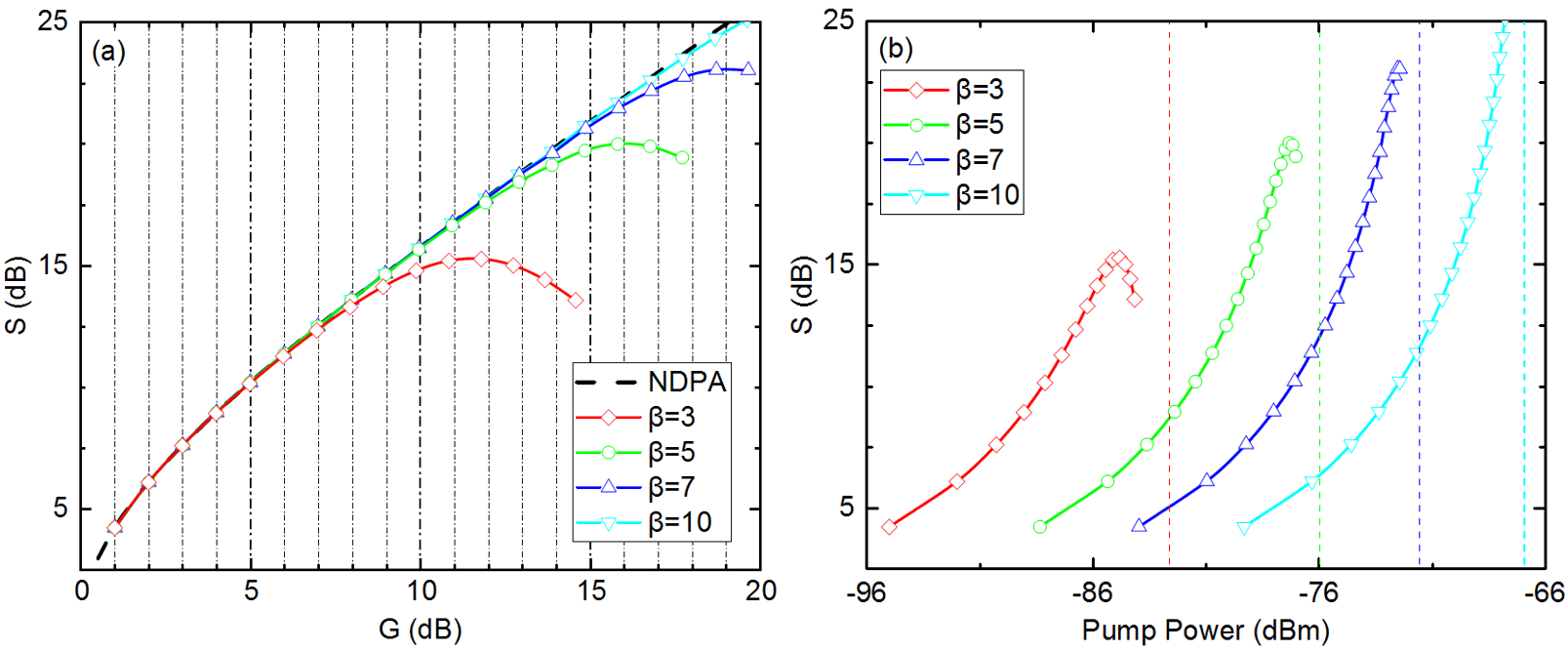}
\caption{Squeezing $S$ as a function of (a) actual gain $G$ or (b) pump power under the Hamiltonian $H_{J1}$. We compare four systems $\beta = 3, 5, 7, 10$ with the ideal NDPA model. Points indicated with markers are actual results for integer values of expected gain $G_0$. The left shift of markers compared to vertical dashed lines in (a) shows difference between actual gain $G$ and expected gain $G_0$. Vertical dashed lines in (b) represent the threshold power of each system.}
\label{fig:beta-gs}
\end{figure*}

We numerically solve the master equation using the BiCGSTAB algorithm \cite{HV90} implemented in MATLAB. The stopping criterion is $10^{-10}$ error in relative residual. Initial guesses and preconditioning, which are additional inputs for the algorithm, are important for convergence and computational efficiency. Initial guesses come from the NDPA model with the stiff pump approximation. We can exactly solve the master equation of the NDPA model, so solutions from the NDPA model serve as initial guesses for the master equations of other systems \cite{SUPP}.

Preconditioning is, loosely speaking, forming an approximate inverse of the problem. For preconditioning, we solve the master equation related to the following Hamiltonian:
\begin{equation} \label{prec-ham}
H = \Delta_a \hat{a}^\dagger \hat{a} + \Delta_b \hat{b}^\dagger \hat{b} + \Delta_c \hat{c}^\dagger \hat{c} + i \epsilon \left( \hat{c}^\dagger - \hat{c} \right).
\end{equation}
Eq. (\ref{me}) with the above Hamiltonian is solvable at a cheap cost, as the equation is block upper diagonal where blocks are defined by constant $a_1, b_1, a_2$, and $b_2$. We use a one-step red-black Gauss-Seidel iteration as the preconditioner for better performance \cite{YS03}. Red-black partitioning is done via the parity of $a_1 - a_2$. We reduced variables, analytically solved the NDPA model, and identified appropriate block structure for preconditioning, which are the main contributions for making the master equation a tractable problem.

It is also possible to compute gain at low signal power. Phase preserving gain is defined as
\begin{equation}
G = 10\, \text{Log}_{10} \left| \frac{\braket{\hat{a}_{out}}}{\braket{\hat{a}_{in}}} \right|^2 .
\end{equation}
If the input field of the signal mode is a coherent state with parameter $\alpha$, this effectively adds $i\sqrt{\kappa_a} \left( \alpha \hat{a}^\dagger - \alpha^* \hat{a} \right)$ to the Hamiltonian. Using a perturbative approach, we write the steady state as $\rho_{ss} (\alpha) = \rho_{ss} + \alpha \rho_1 + \alpha^* \rho_1^\dagger +\cdots$. The steady state equation at order $\alpha$ is $\mathcal{L}[\rho_1]+\sqrt{\kappa_a} [\hat{a}^\dagger, \rho_{ss}] = 0$. $\rho_{ss}$ is the steady state with vacuum input which has elements only in $V_0$, so $[\hat{a}^\dagger, \rho_{ss}]$ has elements only in $V_1$. The same methods for computing initial guesses and preconditioning are applicable to finding $\rho_1$. Gain is obtained from
\begin{equation}
\frac{\braket{\hat{a}_{out}}}{\braket{\hat{a}_{in}}} = \left. \frac{\partial \braket{\hat{a}_{out}}}{\partial \alpha} \right|_{\alpha=0} = \sqrt{\kappa_a}\, \text{Tr} \left( \hat{a} \rho_1 \right)-1,
\end{equation}
using the boundary condition Eq. (\ref{bc}). When setting pump power $\epsilon$ to achieve expected gain $G_0$ in the NDPA model, the actual gain $G$ will differ from $G_0$. We define this difference as \textit{reduction in gain}, $(G-G_0)/G_0$.

\begin{table}[b]
\caption{\label{tab:trunc-lev}
Truncation levels $n$ and $n_c$ used with respect to expected gain $G_0$. The first $n$, $n_c$ levels are used in computation. }
\begin{ruledtabular}
\begin{tabular}{cccccc}
$G_0$ (dB) & 1$\sim$14 & 15$\sim$17 & 18 & 19 & 20  \\  \hline 
$n$ & 20 & 30 & 34 & 36 & 40\\
$n_c$ ($\beta=3,5,7$) & 20 & 20 & 20 & 20 &20\\
$n_c$ ($\beta = 10$) & 30 & 30 & 30 & 30 & 30 
\end{tabular}
\end{ruledtabular}
\end{table}

\begin{figure*}[t]
\includegraphics[width=18cm]{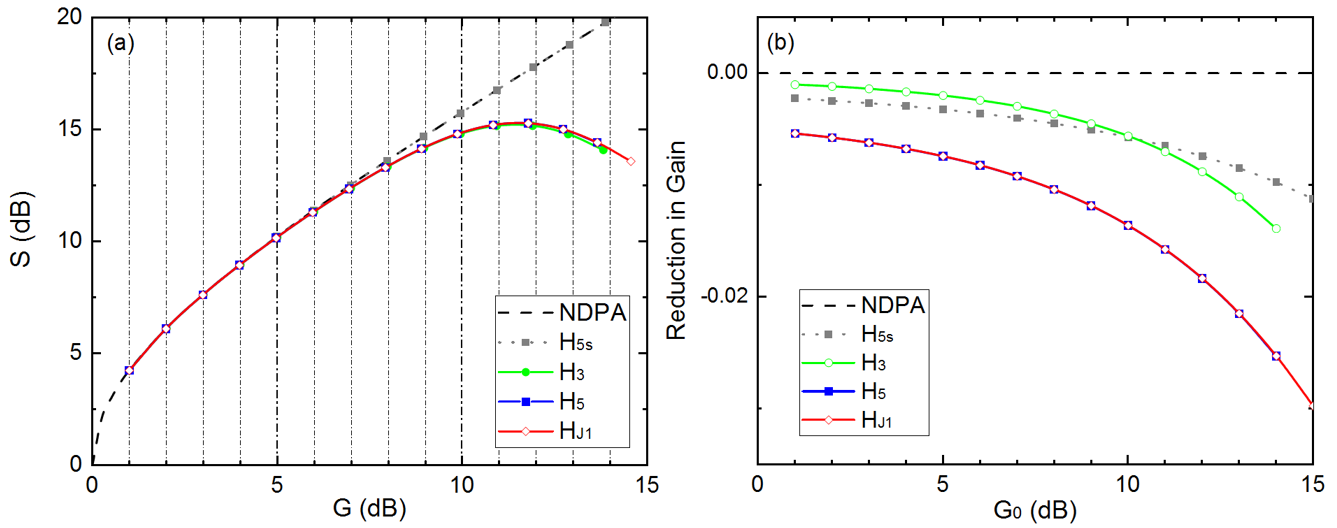}
\caption{(a) Squeezing $S$ as a function of actual gain $G$ under various Hamiltonians with $\beta = 3$. The Hamiltonians are $H_{3s} = H_{NDPA}$, $H_{5s}$, $H_3$, $H_5$, and $H_{J1}$. (b) Reduction in gain $(G-G_0)/G_0$ as a function of expected gain $G_0$ under the various Hamiltonians as in (a). Results of $H_5$ overlap with those of $H_{J1}$. The red lines $H_{J1}$ here correspond to the red lines in Fig. \ref{fig:beta-gs}.}
\label{fig:gs-mod}
\end{figure*}

Here, we set the system parameters as: $\omega_a = 7.5\times2\pi$ GHz, $\omega_b = 5\times 2\pi$ GHz, $\kappa_a =\kappa_b = \kappa_c = 100\times 2\pi$ MHz, $i_c = 1\; \mu\text{A}$ as in Ref. \cite{CL20}, $\Delta_a = \Delta_b =0$, and $\omega_P = \omega_a+\omega_b$. The truncation level $n$ of signal, idler modes and truncation level $n_c$ of pump mode are set as in Table \ref{tab:trunc-lev}. $n$ is determined so that the error in $S$ from truncation is less than 1\% in the NDPA model. The required pump power for same gain increases as $\beta^3$ (see Eq. (\ref{3SoP})), so considering higher levels of the pump mode is necessary when solving problems with high $\beta$. The obtained numerical values were shown to converge as increasing $n$ and $n_c$.  For problems with the stiff pump approximation, we set $n=40$ and directly solve the master equation.

{\textbf{Results.} Fig. \ref{fig:beta-gs}(a) shows $S$ as a function of actual gain $G$ for $\beta = 3,5,7,10$, using the Hamiltonian $H_{J1}$. There is a maximal point for $S$ in terms of $G$, which limits the operation of JRM as an entanglement source. At low gain, squeezing is similar to that of the ideal NDPA model, but at high gain, squeezing starts to deviate from the NDPA model. Deviation occurs at higher gain for systems with higher $\beta$, which corresponds to smaller nonlinearity. We plot the same results with respect to pump power in Fig. \ref{fig:beta-gs}(b). To achieve same squeezing or gain, higher pump power is required for systems with smaller nonlinearity. However, this would heat up the system adding thermal fluctuations or other spurious effects which also limit squeezing. These imply that there will be an optimal $\beta$ for JRM as a squeezing source in realistic systems.

To understand the mechanism of the squeezing limit, we compute $S$ and $G$ with respect to various Hamiltonians, $H_{3s}=H_{NDPA}$, $H_{5s}$, $H_3$, $H_5$, and $H_{J1}$. Fig. \ref{fig:gs-mod} shows $S$ and $(G-G_0)/G_0$ respectively computed from the Hamiltonians with $\beta = 3$. Squeezing limit appears for the full third-order Hamiltonian $H_3$, whereas squeezing from $H_{5s}$ does not deviate from the NDPA model. This shows that the three-wave mixing term $\hat{a} \hat{b} \hat{c}^\dagger$ itself imposes a bound on squeezing. Such effect is not profound in the optical regime using nonlinear crystals, where the main limiting factor is attributed to optical losses \cite{RS17}. However, decrease of squeezing at high pump power was observed experimentally with JRM in Ref. \cite{SB20}. Squeezing is a delicate process, in which various losses, broadened bandwidth of realistic detectors, system instability, and so on can restrict its performance. Excluding these extrinsic factors, our results show that the three-wave mixing interaction $\hat{a} \hat{b} \hat{c}^\dagger$ itself intrinsically limits squeezing, providing fundamental insight to this interaction.

Fig. \ref{fig:gs-mod}(b) shows $(G-G_0)/G_0$ under various Hamiltonians. The decrease in gain from fifth-order terms, $H_{5s}$, is easily understood as the fifth-order terms effectively reduce the three-wave mixing coupling constant depending on mean photon number of each mode. Even for the $H_3$ Hamiltonian, there is a reduction in gain which is comparable to the drop introduced from the fifth-order terms, $H_{5s}$. Again, the $\hat{a} \hat{b} \hat{c}^\dagger$ term itself reduces gain.

\begin{figure*}[t]
\includegraphics[width=18cm]{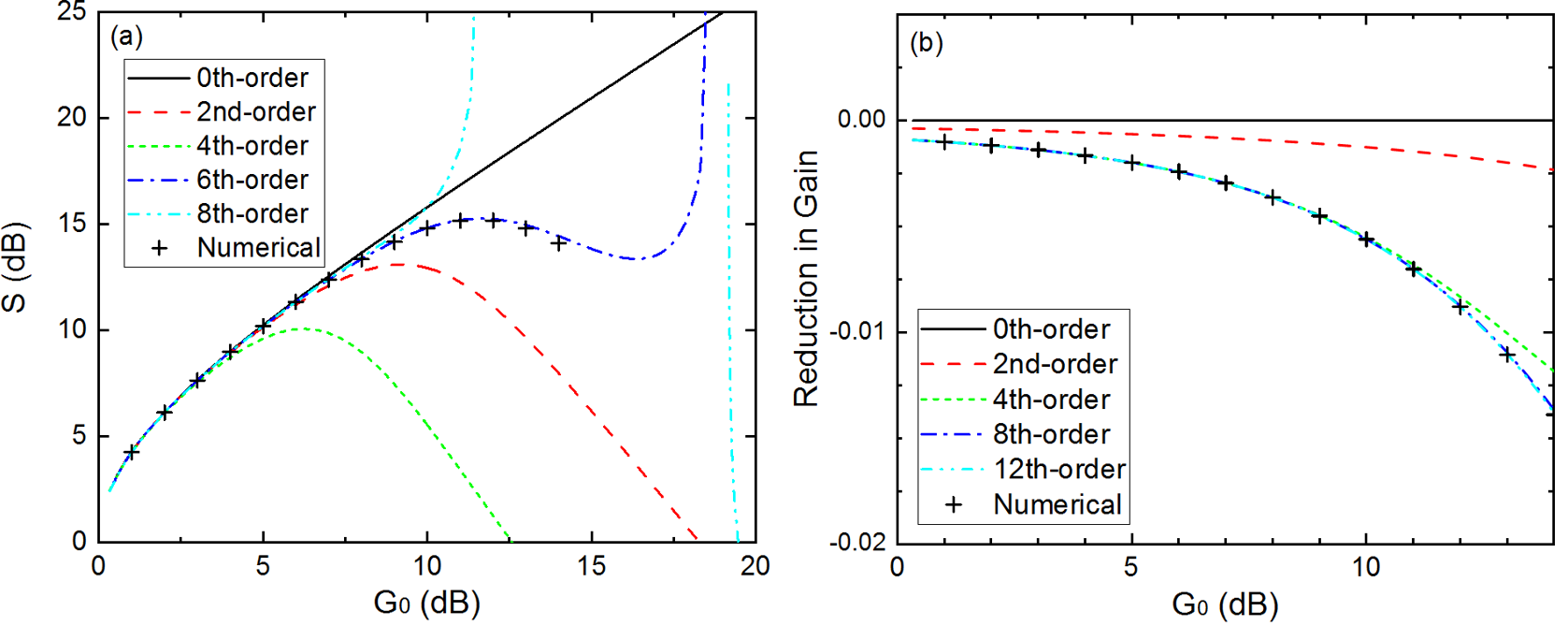}
\caption{(a) Squeezing $S$ and (b) reduction in gain $(G-G_0)/G_0$ as a function of expected gain $G_0$ for the Pad\'e approximants and numerical results under $\beta=3$ and $H_3$. Zeroth-order results and numerical results here correspond to the results of NDPA and $H_3$ in Fig. \ref{fig:gs-mod}, respectively.}
\label{fig:gain-pert}
\end{figure*}

To further verify that the $\hat{a} \hat{b} \hat{c}^\dagger$ term is responsible for altering $S$ and $G$, we use a power series expansion in solving the third-order Hamiltonian $H_3$. Displacing mode $\hat{c}$ by $-\epsilon/\left( \frac{\kappa_c}{2}+i\Delta_c \right)$, the master equation becomes $\dot{\rho} = \mathcal{L}_0 [\rho] + \mathcal{L}_1 [\rho]$ with
\begin{align}
\mathcal{L}_0[\rho] &=-i[H_0, \rho] + \sum_{m=a,b,c} \frac{\kappa_m}{2} \mathcal{D}_{\hat{m}} [\rho],\\
\mathcal{L}_1 [\rho] &= \sqrt{\frac{\omega_a \omega_b \omega_c}{2\beta^3 \omega_J}}[\hat{a} \hat{b} \hat{c}^\dagger - \hat{a}^\dagger \hat{b}^\dagger \hat{c}, \rho],
\end{align}
where $H_0 = H_{NDPA} + \Delta_c \hat{c}^\dagger \hat{c}$. Treating $\mathcal{L}_1$ as a perturbation, the zeroth-order steady state $\rho_0$ is nothing but a two-mode squeezed thermal state in modes $\hat{a}$, $\hat{b}$ and vacuum in mode $\hat{c}$. The actual steady state is $\rho_{ss} = (I+\mathcal{L}_0^{-1} \mathcal{L}_1 )^{-1} [\rho_0]$, so we can write $\rho_{ss}$ as a power series in $\mathcal{L}_1$, provided it converges. The output field moments and gain are computable through $\rho_{ss}$ and $\mathcal{L}^{-1} = (\mathcal{L}_0 + \mathcal{L}_1 )^{-1}$, so we can write these in a power series of $\mathcal{L}_1$. Only even-order terms are nonzero in the power series, as we take the trace when computing output field moments and gain. The power series are then converted to Pad\'e approximants for better accuracy \cite{CB99}.  Fig. \ref{fig:gain-pert} shows $S$ and $(G-G_0)/G_0$ of various approximants and the numerical results under $\beta = 3$ and $H_3$. As $S$ is the logarithm of a small positive value, it is much more prone to approximation errors. The difference of moments can be negative in some cases, giving the divergences for 6th- and 8th-order plots in Fig. \ref{fig:gain-pert}(a). Meanwhile Fig. \ref{fig:gain-pert}(b) shows that the approximants appropriately account the reduction in gain. These show that the signal and idler loss channels from $\hat{a} \hat{b} \hat{c}^\dagger$ are the main reason for the decline in $S$ and $G$.

\textbf{Conclusion.} We numerically solved the master equation of JRM and computed squeezing of the output fields. Restriction to variables that are relevant and appropriate initial guesses and preconditioner enabled us to compare various model Hamiltonians with available resources. Our block Gauss-Seidel preconditioner is based on the physical observation that a strong pump is used. It is parallelizable for high-performance computers and applicable to other quantum systems which use an external pump, hence a plethora of systems can be treated quantum mechanically. We showed that gain at low signal power is less than that expected from the NDPA model and squeezing limit exists. By considering various model Hamiltonians, we found that the three-wave mixing term $\hat{a} \hat{b} \hat{c}^\dagger $ itself bounds squeezing $S$ and also reduces gain. A power series expansion was also computed to support the results. Our results are system independent and apply to general non-degenerate amplifiers that employ three-wave mixing processes.

Results of the full sine potential of JRM are close to those of the fifth-order Hamiltonian. Increasing $\beta$ decreases the overall nonlinearity of the JRM, and the maximum achievable $S$ increases. However, as nonlinearity decreases, a stronger pump is required for same squeezing or gain, and then other effects such as heating from the pump can also limit $S$. A recent study about Josephson traveling wave parametric amplifiers also observed a reduction in squeezing at high gain \cite{ME22}. We suggest that the strong nonlinear interaction itself could cause squeezing limit in other Josephson junction based systems. Our study can be utilized in designing microwave devices for quantum enhanced sensing and measurements.

This work was supported by a grant to Defense-Specialized Project funded by Defense Acquisition Program Administration and Agency for Defense Development.

\newpage
\onecolumngrid
\begin{center}
\textbf{Supplemental Material: Squeezing Limits of the Josephson Ring Modulator\\as a Non-Degenerate Parametric Amplifier}
\end{center}
\section{Quantum Hamiltonian of the JRM}
The classical Hamiltonian of the Josephson Ring Modulator at the Kerr nulling point is given as \cite{CC20}
\begin{equation}
H = \frac{1}{4\phi_0^2} \left(  \frac{2q_a^2}{C_a}  + \frac{2q_b^2}{C_b} + \frac{(C_a+C_b) q_c^2}{2C_a C_b}\right) +\frac{\phi_0^2}{4L_{in}} \left( \varphi_a^2 + \varphi_b^2 +2 \varphi_c^2 \right) - \frac{4\phi_0^2}{L_J} \sin \frac{\varphi_a}{2}  \sin \frac{\varphi_b}{2} \sin \varphi_c.
\end{equation}
The resonance frequencies are
\begin{equation}
\omega_a^2 = \frac{1}{2 C_a L_{in}},\hspace{3mm} \omega_b^2 = \frac{1}{2 C_b L_{in}},\hspace{3mm} \omega_c^2 = \frac{C_a + C_b}{4 C_a C_b L_{in}}.
\end{equation}
There is a relation $\omega_c^2 = \frac{1}{2} \left( \omega_a^2 + \omega_b^2 \right)$. Quantization is done by imposing 
\begin{align}
\hat{\varphi}_a &= \sqrt{\frac{2 \omega_a}{\beta\omega_J}}  \hat{x}_a,\hspace{3mm} \hat{\varphi}_b = \sqrt{\frac{2 \omega_b}{\beta\omega_J}}  \hat{x}_b,\hspace{3mm} \hat{\varphi}_c = \sqrt{\frac{\omega_c}{\beta\omega_J}}  \hat{x}_c,\\
\hat{q}_a &= \hbar \sqrt{\frac{\beta\omega_J}{2 \omega_a}}  \hat{p}_a,\hspace{2mm} \hat{q}_b =\hbar \sqrt{\frac{\beta\omega_J}{2 \omega_b}}  \hat{p}_b,\hspace{2mm} \hat{q}_c =\hbar \sqrt{\frac{\beta\omega_J}{\omega_c}}  \hat{p}_c,
\end{align}
where $\hat{x}_m = \frac{1}{\sqrt{2}} (\hat{m}^\dagger + \hat{m})$, $\hat{p}_m = \frac{i}{\sqrt{2}} ( \hat{m}^\dagger - \hat{m})$ ($m=a,b,c$). Then the Hamiltonian is
\begin{equation}
H_{JRM} = \sum_{m = a,b,c}\hbar \omega_m \hat{m}^\dagger \hat{m}   - 4\hbar\omega_J \sin \frac{\hat{\varphi}_a}{2} \sin \frac{\hat{\varphi}_b}{2} \sin \hat{\varphi}_c .
\end{equation}

We add a pump term to $H_{JRM}$ and move to a rotating frame at input field frequencies $\omega_S$, $\omega_I$, $\omega_P$. Applying the rotating wave approximation (RWA) and linearizing the sine potential gives the third-order Hamiltonian,
\begin{equation}
H_{3} = \sum_{m = a,b,c} \Delta_m \hat{m}^\dagger \hat{m} + i\epsilon \left( \hat{c}^\dagger - \hat{c} \right)  + i\sqrt{\frac{\omega_a \omega_b \omega_c}{2\beta^3 \omega_J}} \left( \hat{a} \hat{b} \hat{c}^\dagger - \hat{a}^\dagger \hat{b}^\dagger \hat{c} \right).
\end{equation}
Adding the next order contribution of the sine potential gives the fifth-order Hamiltonian,
\begin{align} \begin{split}
H_5 &= H_3-i\sqrt{ \frac{\omega_a \omega_b \omega_c}{8 \beta^5 \omega_J^3}} \left[ \frac{\omega_a}{2}  (\hat{a}^\dagger \hat{a}+1) \hat{a} \hat{b} \hat{c}^\dagger   + \frac{\omega_b}{2} (\hat{b}^\dagger \hat{b}+1) \hat{a} \hat{b} \hat{c}^\dagger  + \omega_c \hat{a} \hat{b} \hat{c}^\dagger (\hat{c}^\dagger \hat{c}+1)  \right]+\text{c.c.}.
\end{split} \end{align}
Replacing $\hat{c}$ with $\epsilon /(\frac{\kappa_c}{2} +i\Delta_c)$ in $H_3$, $H_5$ leads to the stiff-pump approximation Hamiltonians,
\begin{align}
H_{3s} &= H_{NDPA}= \sum_{m=a,b} \Delta_m \hat{m}^\dagger \hat{m} + i \left(g \hat{a} \hat{b} - g^* \hat{a}^\dagger \hat{b}^\dagger \right),
\hspace{3mm}g= \sqrt{\frac{\omega_a \omega_b \omega_c}{2\beta^3 \omega_J}} \frac{\epsilon }{(\frac{\kappa_c}{2} -i\Delta_c)},\\
H_{5s} &= H_{3s} - \frac{g}{2\beta\omega_J}\left[ \frac{\omega_a}{2}  (\hat{a}^\dagger \hat{a}+1) \hat{a} \hat{b}  + \frac{\omega_b}{2} (\hat{b}^\dagger \hat{b}+1) \hat{a} \hat{b} + \omega_c \hat{a} \hat{b} \left(\frac{\epsilon^2}{\kappa_c^2/4+\Delta_c^2}+1\right)  \right]+\text{c.c.}.
\end{align}

The explicit form of matrix elements of $\sin \alpha \hat{x}$ is obtained from writing $\sin \alpha \hat{x}$ as a sum of two displacement operators \cite{KC69}. For $n$ and $m$ with different parity,
\begin{equation}
\braket{n|\sin \alpha\hat{x}|m} = e^{-\frac{\alpha^2}{4}}\sqrt{\frac{\min(n,m)!}{\max(n,m)!}} \frac{\alpha}{\sqrt{2}} \left( -\frac{\alpha^2}{2} \right)^{\frac{|n-m|-1}{2}} L_{\min(n,m)}^{|n-m|} \left( \frac{\alpha^2}{2} \right),
\end{equation}
where $L_n^a$ are the generalized Laguerre polynomials. The constant term of $L_n^a$ is $\big(\begin{smallmatrix} n+a\\n\end{smallmatrix}\big)$, so for small fixed $\alpha$, the matrix elements $|\braket{n|\sin\alpha\hat{x}|n+k}|$ decay in $k$ as 
\begin{equation}
|\braket{n|\sin\alpha\hat{x}|n+k}| \propto \sqrt{\frac{n!}{(n+k)!}} \left( \frac{\alpha}{\sqrt{2}} \right)^k \begin{pmatrix} n+k \\ n \end{pmatrix} \propto \sqrt{\frac{(n+k)!}{k!^2}} \left( \frac{\alpha}{\sqrt{2}} \right)^k.
\end{equation}
Retaining only $\hat{a}\hat{b}\hat{c}^\dagger$ like interactions of the JRM sine potential, we obtain the Hamiltonian,
\begin{align}
&H_{J1} = \sum_{m = a,b,c} \Delta_m \hat{m}^\dagger \hat{m} + i\epsilon \left( \hat{c}^\dagger - \hat{c} \right) \nonumber  \\
& +i \sqrt{\frac{\omega_a \omega_b \omega_c}{2\beta^3 \omega_J}} e^{-\frac{\omega_a + \omega_b + 2\omega_c}{8\beta\omega_J}}\sum_{n_a,n_b,n_c} L^1_{n_a} \left( \tfrac{\omega_a}{4\beta\omega_J} \right)  L^1_{n_b} \left( \tfrac{\omega_b}{4\beta\omega_J} \right) L^1_{n_c} \left( \tfrac{\omega_c}{2\beta\omega_J} \right) \frac{\ket{n_a, n_b, n_c+1}\bra{n_a+1, n_b+1, n_c}}{\sqrt{(n_a+1)(n_b+1)(n_c+1)}} \nonumber \\
&+\text{c.c.}.
\end{align}
\section{Solving NDPA via Input-Output Theory}
The ideal non-degenerate parametric amplifier(NDPA) Hamiltonian in a frame rotating at the signal, idler frequencies ($\omega_s, \omega_i$ respectively) is
\begin{equation} 
H_{NDPA} = \Delta_a \hat{a}^\dagger \hat{a} + \Delta_b \hat{b}^\dagger \hat{b} +ig\left( e^{-i\theta} \hat{a} \hat{b} -e^{i\theta} \hat{a}^\dagger \hat{b}^\dagger \right).
\end{equation}
$\Delta_a := \omega_a - \omega_s$ and $\Delta_b := \omega_b - \omega_i$ are detunings of cavity resonances from signal and idler frequencies, and $g>0$ is coupling strength. Defining the Fourier transform as
\begin{equation}
f[\omega] = \int \frac{dt}{\sqrt{2\pi}}\; f(t) e^{i\omega t},
\end{equation}
we can write the equation of motion of operators and input-output relations in frequency domain as
\begin{align}
\begin{bmatrix} \sqrt{\kappa_a}&\\&\sqrt{\kappa_b} \end{bmatrix}
\begin{bmatrix} \hat{a}_{i} [\omega] \\\hat{b}_{i} [-\omega]^\dagger \end{bmatrix}
&= \begin{bmatrix} i(\Delta_a -\omega) + \frac{\kappa_a}{2} & ge^{i\theta} \\ ge^{-i\theta} &  -i(\Delta_b +\omega) + \frac{\kappa_b}{2}  \end{bmatrix} 
\begin{bmatrix} \hat{a}[\omega] \\ \hat{b}[-\omega]^\dagger \end{bmatrix},\\
\begin{bmatrix} \hat{a}_{o} [\omega] \\\hat{b}_{o} [-\omega]^\dagger \end{bmatrix}
&= \begin{bmatrix} \sqrt{\kappa_a}&\\&\sqrt{\kappa_b} \end{bmatrix}
 \begin{bmatrix} \hat{a} [\omega] \\\hat{b} [-\omega]^\dagger \end{bmatrix} 
-  \begin{bmatrix} \hat{a}_{i} [\omega] \\\hat{b}_{i} [-\omega]^\dagger \end{bmatrix}.
\end{align}
$\kappa_a$ and $\kappa_b$ are the mode decay rates which describe dissipation. Solving the equations give
\begin{align}
\begin{bmatrix} \hat{a}_{o} [\omega] \\\hat{b}_{o} [-\omega]^\dagger \end{bmatrix}
&= \left(
\begin{bmatrix} \sqrt{\kappa_a}&\\&\sqrt{\kappa_b} \end{bmatrix}
\begin{bmatrix} i(\Delta_a -\omega) + \frac{\kappa_a}{2} & ge^{i\theta} \\ ge^{-i\theta} &  -i(\Delta_b +\omega) + \frac{\kappa_b}{2}  \end{bmatrix} ^{-1}
\begin{bmatrix} \sqrt{\kappa_a}&\\&\sqrt{\kappa_b} \end{bmatrix} -1 \right) 
\begin{bmatrix} \hat{a}_{i} [\omega] \\\hat{b}_{i} [-\omega]^\dagger \end{bmatrix}\\
&=S[\omega] \begin{bmatrix} \hat{a}_{i} [\omega] \\\hat{b}_{i} [-\omega]^\dagger \end{bmatrix}. \label{Smat}
\end{align}
Defining susceptibilities as
\begin{equation}
\chi_a^{-1} := \frac{\kappa_a}{2} +i(\Delta_a-\omega),\hspace{3mm} \chi_b^{-1} := \frac{\kappa_b}{2} +i(\Delta_b+\omega),
\end{equation}
we can write
\begin{align}
S_{11} &= \frac{g^2+ (\chi_a^*)^{-1} (\chi_b^*)^{-1}}{\chi_a^{-1} (\chi_b^*)^{-1}-g^2},\; S_{12} = \frac{-g \sqrt{\kappa_a \kappa_b} e^{i\theta}} {\chi_a^{-1} (\chi_b^*)^{-1}-g^2},\\
S_{21} &= \frac{-g \sqrt{\kappa_a \kappa_b} e^{-i\theta}}{\chi_a^{-1} (\chi_b^*)^{-1}-g^2},\; S_{22} = \frac{g^2+ \chi_a^{-1} \chi_b^{-1}} {\chi_a^{-1} (\chi_b^*)^{-1}-g^2}.
\end{align}
The low power gain is $|S_{11}|^2$. The nonzero second-order moments are
\begin{equation}  \label{ndpa-mom}
\braket{\hat{a}_{o}[\omega]^\dagger \hat{a}_{o}[\omega]} = |S_{12}|^2,\;
\braket{\hat{b}_{o}[-\omega]^\dagger \hat{b}_{o}[-\omega]} = |S_{21}|^2,\;
\braket{\hat{a}_{o}[\omega] \hat{b}_{o}[-\omega]} = S_{11}S^*_{21}.
\end{equation}
We have omitted a factor of $\delta(0)$ which comes from $\left[\hat{a}_o[\omega_1], \hat{a}_o [\omega_2]^\dagger \right] = \delta(\omega_1-\omega_2)$. We also have omitted a factor of $\delta(0)$ in the numerical results of the main text, which corresponds to considering an ideal detector of a very narrow bandwidth.

\section{Solving NDPA via Quantum Regression Theorem}
We solve the ideal NDPA by the methods described in the main text. The calculations are complicated than the above scattering matrix approach in frequency domain. The purpose of doing these calculations is two-fold: verify that this approach is equivalent to the scattering matrix approach and provide appropriate initial guesses for the numerical methods used in solving the full Hamiltonian. The master equation describing the dynamics of the density operator is
\begin{equation} \label{ndpa-me}
\dot{\rho} = -i[H_{NDPA}, \rho] + \frac{\kappa_a}{2} \mathcal{D}_{\hat{a}} [\rho] + \frac{\kappa_b}{2} \mathcal{D}_{\hat{b}} [\rho]=:\mathcal{L}[\rho],
\end{equation}
where $\mathcal{D}_{\hat{a}} [\rho] := 2 \hat{a} \rho \hat{a}^\dagger - \rho \hat{a}^\dagger \hat{a} - \hat{a}^\dagger \hat{a} \rho$ gives dissipation and $\kappa_a, \kappa_b$ are the mode decay rates.

\subsection{Steady State and Eigen-modes of the NDPA Master Equation}
Inverting $\mathcal{L}$ amounts to finding all eigen-modes and eigenvalues of $\mathcal{L}$, where the eigen-mode with eigenvalue 0 is the steady state and other eigen-modes are exponentially damped in time. It is possible to write eigen-modes in terms of characteristic functions, defined as
\begin{equation}
\chi_{\rho}(\alpha, \beta) = \text{Tr} \left( \hat{D}(\alpha, \beta) \rho \right),
\end{equation}
where $\hat{D}(\alpha, \beta) = e^{\alpha \hat{a}^\dagger +\beta \hat{b}^\dagger - \text{c.c.}}$ is the standard displacement operator. The master equation (\ref{ndpa-me}) can be formulated for the characteristic function as
\begin{align} \label{ndpa-mechar}
\begin{split}
\dot{\chi}&= -i\Delta_a \left[ -\alpha \partial_\alpha + \alpha^* {\partial}_{\alpha^*} \right]\chi -i\Delta_b \left [-\beta \partial_\beta + \beta^* {\partial}_{\beta^*} \right]\chi\\
&\hspace{5mm} +ge^{-i\theta} \left[\beta {\partial}_{\alpha^*} + \alpha {\partial}_{\beta^*} \right] \chi +ge^{i\theta} \left[\beta^* \partial_\alpha + \alpha^* \partial_\beta \right] \chi \\
&\hspace{5mm} +  \frac{\kappa_a}{2} \left[- \alpha^* {\partial}_{\alpha^*} -\alpha \partial_\alpha - |\alpha|^2  \right] \chi 
+\frac{\kappa_b}{2} \left[- \beta^* {\partial}_{\beta^*} -\beta \partial_\beta - |\beta|^2  \right] \chi.
\end{split}
\end{align}

One can show that the steady state solution to the above equation is a Gaussian function.
\begin{equation}
\chi_{ss} (\alpha, \beta) = e^{-q^\dagger \sigma q/2},\;
\sigma = \begin{bmatrix}
s_1 &&& s_2 \\ &s_1&s_3 & \\ &s_2&s_4&\\ s_3 &&& s_4
\end{bmatrix}.
\end{equation}
The coordinates $q$ are $q = (\alpha, -\alpha^*, \beta, -\beta^*)^t$. This state is a two mode squeezed thermal state. Entries of $\sigma$ give the symmetrized moments of the steady state.
\begin{align}
s_1 = \braket{\hat{a}^\dagger \hat{a} }+\tfrac{1}{2},\;s_2 = \braket{\hat{a} \hat{b}},\; s_3 = \braket{\hat{a}^\dagger \hat{b}^\dagger},\; s_4 = \braket{\hat{b}^\dagger \hat{b} }+\tfrac{1}{2}
\end{align}
Specific values of $s_1,s_2, s_3, s_4$ are
\begin{align}
s_1 &= \frac{\kappa_a \kappa_b (\kappa^2 + 4\Delta^2)+4g^2 (-\kappa_a^2+\kappa_b^2)}{2( \kappa^2(\kappa_a \kappa_b - 4g^2)+ 4\kappa_a \kappa_b \Delta^2)},\\
s_2 &= \frac{-2\kappa_a \kappa_b (\kappa -2i \Delta) g e^{i\theta}}{ \kappa^2(\kappa_a \kappa_b - 4g^2)+ 4\kappa_a \kappa_b \Delta^2},\\
s_3 &= \frac{-2\kappa_a \kappa_b (\kappa +2i \Delta) g e^{-i\theta}}{ \kappa^2(\kappa_a \kappa_b - 4g^2)+ 4\kappa_a \kappa_b \Delta^2},\\
s_4 &= \frac{\kappa_a \kappa_b (\kappa^2 + 4\Delta^2)+4g^2 (\kappa_a^2-\kappa_b^2)}{2( \kappa^2(\kappa_a \kappa_b - 4g^2)+ 4\kappa_a \kappa_b \Delta^2)}.
\end{align}
We introduced the total loss $\kappa := \kappa_a + \kappa_b$ and total detuning $\Delta := \Delta_a + \Delta_b$. The number state expansion of this state contains terms only of the form
\begin{equation}
\ket{n+\ell,n}\bra{m+\ell, m}, \hspace{3mm} \ket{n,n+\ell}\bra{m,m+\ell}.
\end{equation}
The coefficient of $\ket{n+\ell,n}\bra{m+\ell, m}$ is
\begin{equation} \label{ndpa-ssnum1}
4 \frac{\kappa_a\kappa_b (\kappa^2+4\Delta^2)-4\kappa^2 g^2}{4\kappa_a \kappa_b (\kappa^2+4\Delta^2-4g^2)} \sqrt{\frac{(n+\ell)!(m+\ell)!}{n!m!\ell!^2}}  a^\ell c^n d^m  \;_2 F_1 \left(-n,-m;\ell+1;\frac{4g^2}{\kappa^2+4\Delta^2} \right), 
\end{equation}
and the coefficient of $\ket{n,n+\ell}\bra{m,m+\ell}$ is
\begin{equation} \label{ndpa-ssnum2}
4 \frac{\kappa_a\kappa_b (\kappa^2+4\Delta^2)-4\kappa^2 g^2}{4\kappa_a \kappa_b (\kappa^2+4\Delta^2-4g^2)} \sqrt{\frac{(n+\ell)!(m+\ell)!}{n!m! \ell!^2}}  b^\ell c^n d^m  \;_2 F_1 \left(-n,-m;\ell+1;\frac{4g^2}{\kappa^2+4\Delta^2} \right).
\end{equation}
$\,_2 F_1$ is the hypergeometric function. $a, b, c, d$ are numbers given as
\begin{align} \begin{split}
a &= \frac{4\kappa_b g^2/\kappa_a}{\kappa^2+4\Delta^2-4g^2},\;
b = \frac{4\kappa_a g^2/\kappa_b}{ \kappa^2+4\Delta^2-4g^2},\\
c &= \frac{-2(\kappa+2i\Delta)ge^{-i\theta}}{\kappa^2+4\Delta^2-4g^2},\;
d = \frac{-2(\kappa-2i\Delta)ge^{i\theta}}{\kappa^2+4\Delta^2-4g^2}.
\end{split} \end{align}

The other eigen-modes of (\ref{ndpa-mechar}) can be found from the steady state. For some function $f$, the eigenvalue problem of $\chi = f \chi_{ss}$ can be written in terms of $f$ as
\begin{align} \label{ndpa-eigp} \begin{split}
\lambda f&= -i\Delta_a \left[ -\alpha \partial_\alpha + \alpha^* {\partial}_{\alpha^*} \right]f -i\Delta_b \left [-\beta \partial_\beta + \beta^* {\partial}_{\beta^*} \right]f\\
&\hspace{2mm} +ge^{-i\theta} \left[\beta {\partial}_{\alpha^*} + \alpha {\partial}_{\beta^*} \right] f +ge^{i\theta} \left[\beta^* \partial_\alpha + \alpha^* \partial_\beta \right] f \\
&\hspace{2mm} +  \frac{\kappa_a}{2} \left[- \alpha^* {\partial}_{\alpha^*} -\alpha \partial_\alpha\right] f + \frac{\kappa_b}{2} \left[- \beta^* {\partial}_{\beta^*} -\beta \partial_\beta  \right] f.
\end{split} \end{align}
If $f$ is a homogeneous polynomial in $\alpha, \alpha^*, \beta, \beta^*$, then the right-hand side of (\ref{ndpa-eigp}) is a homoegnous polynomial of same degree, hence the eigenfunctions $f$ will be homogeneous polynomials. Also the variable $\alpha$ is coupled only with $\beta^*$, so $f$ will be a product of two polynomials, one with variables $\alpha, \beta^*$ and the other with variables $\alpha^*, \beta$. This reflects the fact that the NDPA mixes the operator $\hat{a}$ only with $\hat{b}^\dagger$.

The degree-1 solutions of (\ref{ndpa-eigp}) are the building blocks of general eigen-modes. The solutions in variables $\alpha, \beta^*$ have eigenvalues
\begin{equation}
\lambda_{\pm} = \frac{1}{2} \left(-(\chi_a^*)^{-1} -\chi_b^{-1}\pm\sqrt{(-(\chi_a^*)^{-1} +\chi_b^{-1})^2+4g^2} \right),
\end{equation}
with $\chi_m^{-1} = \frac{\kappa_m}{2} +i\Delta_m$ $(m=a,b)$. The corresponding eigenfunctions are
\begin{align}
f_+ &= \xi \alpha +2g e^{i\theta} \beta^*, \hspace{3mm} f_- = \xi \beta^* -2g e^{-i\theta} \alpha, \\
\xi &= -(\chi_a^*)^{-1} + \chi_b^{-1} +\sqrt{(-(\chi^*_a)^{-1} + \chi_b^{-1} )^2 +4g^2}.
\end{align}
The degree-1 eigenvalues, eigenfunctions of variables $\alpha^*, \beta$ are simply given by complex conjugates, $\lambda^*_+, \lambda^*_-, f^*_+, f^*_-$. General eigen-modes of (\ref{ndpa-eigp}) are products of these functions, 
\begin{equation}
f_{nm\ell k} = f_+^n {f}_+^{*m} f_-^{*\ell} f_-^k, \hspace{3mm} \lambda_{nm\ell k} = n \lambda_+ + m \lambda^*_+ + \ell \lambda^*_- + k \lambda_-.
\end{equation}
Then characteristic functions of the eigen-modes of the master equation are $\chi_{nm\ell k} = f_{nm\ell k} \chi_{ss}$. Finally, we note that multiplication by $\alpha, \alpha^*$ on $\chi$ corresponds to $[\rho, \hat{a}]$, $[\rho, \hat{a}^\dagger]$ respectively, so with (\ref{ndpa-ssnum1}), (\ref{ndpa-ssnum2}) it is possible to compute the number basis representations of these states. In the following, we refrain from explicitly working with eigen-modes and instead use the fact that $\mathcal{L}$ preserves the degree of $f$ when acting on a state with characteristic function $\chi = f \chi_{ss}$.

\subsection{Output Field Moments of NDPA}
We compute $\braket{\hat{a}_o[\omega] \hat{b}_o[-\omega]}$ using the quantum regression theorm. The integral over $dt_1 dt_2$ is split over time orderings and $\tau = |t_1 - t_2|$.
\begin{align}
\braket{\hat{a}_o[\omega_1] \hat{b}_o[\omega_2]} &= \frac{1}{2\pi} \int dt_1 dt_2\; \braket{\hat{a}_o(t_1) \hat{b}_o(t_2)} e^{i\omega_1 t_1+i \omega_2 t_2}\\
&=\frac{1}{2\pi} \left(\int_{t_1>t_2}  + \int_{t_1<t_2} \right) \text{Tr} \left[ \hat{a}_o (t_1) \hat{a}_o (t_2) \rho_{ss} \right] e^{i\omega_1 t_1 + i\omega_2 t_2}\\
&= \delta(\omega_1+\omega_2) \int_0^\infty d\tau\; \text{Tr} \left[ \hat{a}_o (\tau) \hat{b}_o \rho_{ss} \right] e^{i\omega_1 \tau} +\text{Tr} \left[ \hat{b}_o (\tau) \hat{a}_o \rho_{ss} \right] e^{i\omega_2 \tau}\\
&= \delta(\omega_1+\omega_2)\sqrt{\kappa_a \kappa_b} \int_0^\infty d\tau\; \text{Tr} \left[ \hat{a} e^{\mathcal{L}\tau} [\hat{b} \rho_{ss}] \right] e^{i\omega_1 \tau} +\text{Tr} \left[ \hat{b} e^{\mathcal{L}\tau} [\hat{a} \rho_{ss}] \right] e^{i\omega_2 \tau} \label{use-io}\\
&=\delta(\omega_1+\omega_2)\sqrt{\kappa_a \kappa_b} \left( \text{Tr} \left[ \hat{a} \frac{-1}{\mathcal{L}+i\omega_1} [\hat{b} \rho_{ss}] \right]  +\text{Tr} \left[ \hat{b} \frac{-1}{\mathcal{L}+i\omega_2} [\hat{a} \rho_{ss}] \right]  \right).
\end{align}
We have used the input-output relation $\hat{a}_o = \sqrt{\kappa_a} \hat{a}$ in (\ref{use-io}) which is valid under trace when the input field is vacuum. The characteristic function of $\hat{a} \rho_{ss}$ is
\begin{equation}
\left( - \partial_{\alpha^*}-\frac{\alpha}{2} \right) \chi_{ss} = \begin{bmatrix} \alpha \chi_{ss} & \beta^*\chi_{ss} \end{bmatrix} \begin{bmatrix} s_1-\frac{1}{2}\\ -s_2 \end{bmatrix}.
\end{equation}
Then the characteristic function of $\frac{-1}{\mathcal{L}+i\omega_2}[\hat{a} \rho_{ss}]$ is
\begin{equation} \label{Linv-arho}
\begin{bmatrix} \alpha \chi_{ss} & \beta^*\chi_{ss} \end{bmatrix} 
\left(-\begin{bmatrix}i\Delta_a - \frac{\kappa_a}{2} & ge^{-i\theta} \\ ge^{i\theta} & -i\Delta_b - \frac{\kappa_b}{2} \end{bmatrix}-i\omega_2 \right)^{-1}
 \begin{bmatrix} s_1-\frac{1}{2}\\ -s_2 \end{bmatrix}.
\end{equation}
This serves as an initial guess for the numerical computations of the full Hamiltonian. The trace of $\hat{b} \frac{-1}{\mathcal{L}+i\omega_2}[\hat{a} \rho_{ss}]$ is the negative of the coefficient of $\beta^*$ in (\ref{Linv-arho}). Hence we have
\begin{equation}
\text{Tr} \left[ \hat{b} \frac{-1}{\mathcal{L}+i\omega_2} [\hat{a} \rho_{ss}] \right]  = \begin{bmatrix} 0 & 1 \end{bmatrix} 
\left(\begin{bmatrix}i\Delta_a - \frac{\kappa_a}{2} & ge^{-i\theta} \\ ge^{i\theta} & -i\Delta_b - \frac{\kappa_b}{2} \end{bmatrix}+i\omega_2 \right)^{-1} \begin{bmatrix} s_1-\frac{1}{2} \\ -s_2 \end{bmatrix}.
\end{equation}
Similarly, we can compute
\begin{equation}
\text{Tr} \left[ \hat{a} \frac{-1}{\mathcal{L}+i\omega_1} [\hat{b} \rho_{ss}] \right]  = \begin{bmatrix} 1 & 0 \end{bmatrix} 
\left(\begin{bmatrix}-i\Delta_a - \frac{\kappa_a}{2} & ge^{-i\theta} \\ ge^{i\theta} & i\Delta_b - \frac{\kappa_b}{2} \end{bmatrix}+i\omega_1 \right)^{-1}
 \begin{bmatrix} -s_2 \\ s_4-\frac{1}{2} \end{bmatrix}.
\end{equation}
Combining these results yields the same expression as (\ref{ndpa-mom}). Other moments are computed in the same way.

\subsection{Gain at Low Signal Power of NDPA}
As written in the main text, the gain at low signal power is computed from
\begin{equation}
\frac{\braket{\hat{a}_o}}{\braket{\hat{a}_i}} = \sqrt{\kappa_a} \text{Tr} \left[ \hat{a} \rho_{1} \right]-1,\; G  = 10 \text{ Log}_{10} \left| \frac{\braket{\hat{a}_o}}{\braket{\hat{a}_i}} \right|^2
\end{equation}
where $\rho_1$ is the solution to $\mathcal{L}[\rho_1] = \sqrt{\kappa_a} [\rho_{ss},\hat{a}^\dagger]$. The characteristic function of $[\rho_{ss},\hat{a}^\dagger]$ is just $\alpha^*\chi_{ss}$, so the characteristic function of $\rho_1$ is given as
\begin{equation}
\sqrt{\kappa_a} \begin{bmatrix} \alpha^* \chi_{ss} & \beta \chi_{ss} \end{bmatrix}
\begin{bmatrix}-i\Delta_a - \frac{\kappa_a}{2} & ge^{-i\theta} \\ ge^{i\theta} & i\Delta_b - \frac{\kappa_b}{2} \end{bmatrix}^{-1}
\begin{bmatrix}1\\0 \end{bmatrix} .
\end{equation}
Then we obtain the result of (\ref{Smat}):
\begin{equation}
\frac{\braket{\hat{a}_o}}{\braket{\hat{a}_i}} = \kappa_a \begin{bmatrix}1&0 \end{bmatrix} 
\begin{bmatrix}i\Delta_a + \frac{\kappa_a}{2} & -ge^{-i\theta} \\ -ge^{i\theta} & -i\Delta_b + \frac{\kappa_b}{2} \end{bmatrix}^{-1}
 \begin{bmatrix} 1 \\ 0 \end{bmatrix}-1 = S_{11}[0].
\end{equation}

\end{document}